\documentclass[12pt,a4paper]{article}
\usepackage{amssymb}
\usepackage{amsfonts}
\usepackage[english]{babel}
\usepackage{graphicx}
\usepackage{appendix}

\newcommand{\bean}{\begin{eqnarray*}}
\newcommand{\eean}{\end{eqnarray*}}
\newcommand{\ba}{\begin{array}}
\newcommand{\ea}{\end{array}}
\newcommand{\be}{\begin{equation}}
\newcommand{\ee}{\end{equation}}
\newcommand{\bea}{\begin{eqnarray}}
\newcommand{\eea}{\end{eqnarray}}
\newcommand{\pa}{\partial}
\newcommand{\la}{\lambda}

\newcommand{\no}{\nonumber}
\newcommand{\om}{\omega}

\newcommand{\wt}{\widetilde}
\newcommand{\qd}{\quad}

\begin{document}

\title
{Real Line Solitons  of  the BKP Equation}
\author{
 Jen-Hsu Chang \\Graduate School of National Defense, \\
 National Defense University, \\
 Tauyuan City,  335009, Taiwan }

\date{}

\maketitle
\begin{abstract}
The solitons solution of BKP equation can be constructed by the Pfaffian structure. Then one investigates the real line solitons  structure of BKP equation using the totally non-negative Grassmannian. Especially, the N-soliton solution is studied and its self-dual Tau function is obtained. Also, one can construct the  totally non-negative Grassmannian of the Sawada-Kotera equation for its real line solitons. 

\end{abstract}
\newpage

\section{Introduction} 
\indent The BKP equation \cite{da, dj} or the 2+1 Sawada-Kotera equation \cite{ji} 
\be 
(9 \phi_t-5\phi _{xxy}+\phi_{xxxxx}-15 \phi_x \phi_y+15 \phi_{x}\phi_{xxx}+15\phi_x^3)_x-5 \phi_{yy}=0 \label{bkp}
\ee
is obtained form the reduction of B-type in the KP hierarchy under the orthogonal type transformation group for the KP equation.  It can also be obtained by the Kupershmidt reduction \cite{ku} or the hierarchy defined
on integrable 2D Schrodinger operators \cite{kr}. The complex line solitons  solutions of the BKP equation (\ref{bkp})  are constructed by the vertex operators and the Clifford algebra of free fermions \cite{dj} ($\tau$-function theory), the Pfaffian structure from the Hirota Bilinear form  \cite{hi, rh},  and the gauge transformation between the Lax operators \cite{he}.  In \cite{kv, nim, ni, or, xi}, the rational solutions are established by expressing the $\tau$-functions of the BKP hierarchy as  the linear combinations of the Schur Q-functions (polynomials) over the Pfaffian coefficients defined on  partitions with distinct parts. Also, the $\tau$-functions of the BKP hierarchy can be obtained as the partition function or the matrix integrals \cite{hu, os} to be used in studying the Pfaffian point process \cite{wa}. The non-communicative case of BKP equation and its $\tau$-function are  investigated in \cite{de}. 

On the other hand, the resonant interaction plays a fundamental  role in multi-dimensional wave phenomenon. The resonances of line solitons of KP-(II) equation
\be    \pa_x (-4 u_t+u_{xxx}+6uu_x)+ 3u_{yy}=0  \label{kp} \ee
has attracted much attractions using the totally non-negative Grassmannians \cite{bc, ko1, ko3}, that is, those points of the real Grassmannian whose Plucker coordinates are all non-negative.  For the KP-(II) equation case, the  $\tau$-function  is described by the Wroskian form with respect to $x$ obtained from the Hirota Bilinear form \cite{hi}. Basically, the interaction of KP equation comes from the "X" -shaped O-type soliton (original soliton) due to the angle of intersection smaller than certain critical value depending on the amplitude of the solitons \cite{ko2, mi}.  In this critical angle, the two line-solitons of the O-type solution interact resonantly, and a third line soliton is created to make a "Y" -shaped wave form. It turned out that this "Y" -shaped wave form is also a solution of the KP equation (\ref{kp}). Inspired by the success of 
the totally non-negative Grassmannians applied on the resonances of the KP equation and the coupled KP equation \cite{km},  it is natural to  consider the BKP equation (\ref{bkp}) similarly.

We start with the KP hierarchy. To recall some definitions of hierarchy, we introduce the theory of pseudo-differential operators \cite{dc}. The Lax operator of the KP hierarchy is defined with 
\[L=\pa^{-1}+ u_1 \pa+u_2 \pa^2+ u_3 \pa^3+\cdots ,\]
where $ \pa =\frac{\pa}{\pa x}$ and $u_i$, $i=1,2,3, \cdots, $  are smooth functions of  $t_1=x, t_2, t_3,t_4, \cdots.$. The positive power and the negative power $\pa^n$ is defined by the Leibniz rule 
\be \pa^n f=\sum_0^{\infty} {n \choose k}(\pa^k f) \pa^{n-k}, \label{co} \ee
where ${n \choose k}$ stands for the binomial coefficients 
\[{n \choose k} =\frac{n(n-1)(n-2) \cdots (n-k+1)}{k!}, \mbox{n being any integer} \]
For example,
\bean 
\pa^{-1} f &=& f\pa^{-1}- (\pa f) \pa^{-2} +(\pa^2 f) \pa^{-3} - (\pa^3 f) \pa^{-4} + \cdots  \\
\pa^{-2} f &=& f\pa^{-2}-2 (\pa f) \pa^{-3} +3(\pa^2 f) \pa^{-4} - 4(\pa^3) \pa^{-5} + \cdots.
\eean
The KP hierarchy is given by 
\be \frac{\pa L}{\pa t_n}=[ B_n, L], n=1,2 ,3 \cdots, \label{hi}  \ee
where $B_n=(L^n) _{ \geq 0} $  and   the subscript  $ \geq 0$ means the pseudo-differential $L^n$ is projected  to a  differential operator  and $[A,B]=AB-BA$, the commutator. Then from the zero-curvature condition $ \pa_{t_2} B_3-\pa_{t_3} B_2= [B_3, B_2]$,  letting $u=u_1(x,y,t) $, $t_2=y$, and $t_3=t$, one obtains the Kadomtsev-Petviashvili  (KP-II) equation (\ref{kp}).  The KP-(II) equation (\ref{kp}) also describes the shallow water wave and $u$ is the height of water wave \cite{ko1}. \\
\indent To introduce the BKP hierarchy, we have the Kupershmidt reduction on the Lax operator  $L$ \cite {ku} :
\be L^*=-\pa L \pa^{-1}, \label{ku} \ee
where $L^* $ is the adjoint operator of L define by 
\[ L^*= - \pa^{-1}-   \pa u_1+ \pa^2 u_2 -   \pa^3u_3+ \cdots + (- 1)^n  \pa^n u_n+ \cdots .\]
From (\ref{ku}) and the Leibniz rule (\ref{co}),  a simple calculation can obtain 
\bea u_2 &=& -u_{1x}, \no \\
  u_4 &=& -2 u_{3x}+u_{1xxx} \no \\
	& \vdots & \no \\
	u_{2n} &=&\alpha_n u_{2n-1}^{(1)}+ \alpha_{n-1} u_{2n-3}^{(3)} + \alpha_{n-2} u_{2n-5}^{(5)}+\cdots + \alpha_1 u_{1}^{(2n-1)}, \label{re} 
	  \eea 
where $\alpha_i,  i=1,2 ,3, \cdots, n  $ are constants and $u_i^{(m)}$ is  the differential order of $u_i$ with respect to $x$. For example, $u_{2n-3}^{(3)}=u_{(2n-3)xxx}$. Then the KP hierarchy will reduce to the odd flow 
\be \frac{\pa L}{\pa t_{2n+1}}=[ B_{2n+1}, L], n=1, 2 ,3 \cdots, \label{ck}  \ee
The hierarchy defined by (\ref{ku}) and (\ref{ck}) is the BKP hierarchy. We remark here that the reduction (\ref{ku}) or (\ref{re}) is equivalent to the zero constant-level term for any $B_{2n+1}$ \cite{da}. Similarly, from the zero curvature condition $ \pa_{t_3} B_5-\pa_{t_5} B_3= [B_5, B_3]$, one obtains the BKP equation or the (2+1) Sawada-Kotera equation \cite{de} , introducing potential $\phi $ and $u_1=\phi_x, y=t_3, t=t_5$, 
we have the BKP equation (\ref{bkp}). When $ \phi $ is independent of $y$ (or $\phi_{y}=0$), one has the Sawada-Kotera equation \cite{kk} . The Hirota equation of the BKP equation (\ref{bkp} ) is 
\be (D_x^6-5D_x^3 D_y -5 D_y^2+9 D_x D_t) \tau \circ \tau=0,  \label{hr} \ee
where the Hirota derivative is defined as \cite{hi} 
\[D_t^m D_x^n f(t,x) \circ g(t,x) =\frac{\pa^m} {\pa a^m} \frac{\pa^n} {\pa b^n} f(t+a,x+b) g(t-a,x-b) |_{a=0, b=0} \] 
and  $ \phi=2 (\ln \tau)_x$ or  $ u_1=2 (\ln \tau)_{xx}.$  The N-solitons solution of the BKP equation  (\ref{bkp}) or  (\ref{hr})  is constructed by the Pfaffian structure as follows \cite{rh}. 
\begin{itemize}
\item Choose functions $E_i (x,y,t) $ such that 
\be \frac {\pa E_i }{\pa t}= \frac {\pa^5 E_i}{\pa x^5}, \quad \frac {\pa E_i }{\pa y}= \frac {\pa^3 E_i}{\pa x^3}, \label{li} \ee
where $i=1,2,3, \cdots 2N. $ 
\item Define the skew product 
\be W_{i,j}= \int_{-\infty}^x [D_x (E_i \circ E_j) ] dx =\int_{-\infty}^x (E_{ix} E_j- E_{jx} E_i)  dx. \label{sk} \ee
 Notice that $W_{i,j}= -W_{j,i}, W_{i,i}=0$. 
\item Then 
\bea \tau (x,y,t) &=& Pf(E_1, E_2, E_3, \cdots, E_{2N} ) \no \\
&=& \sum_{\sigma}\epsilon(\sigma)W_{\sigma_1, \sigma_2}W_{\sigma_3
, \sigma_4} \cdots W_{\sigma_{2N-1},  \sigma_{2N}},  \label{pf1} \eea
where $Pf$ is the Pfaffian of $E_1, E_2, \cdots , E_{2N}$ and $ \epsilon(\sigma)$  is the sign function of the permutation $\sigma$; moreover,  
\[ \sigma_1 < \sigma_3< \sigma_5 \cdots < \sigma_{2N-1}, \quad  \sigma_1 < \sigma_2 ,   \sigma_3 < \sigma_4, \cdots,      \sigma_{2N-1} < \sigma_{2N}  .   \] 
\end{itemize}
If we take 
\[E_i(x,y,t)=e^{p_i x+ p_i^3 t+ p_i^5 t+ \eta_i}, \]
where $p_i$ and $\eta_i$(phase) are constants,  and then $W_{i,j}=\frac{p_i-p_j}{p_i+p_j} E_i E_j$. Thus, using the  the Schur identity  \cite{jh1, is},
 \be Pf(\frac{p_i-p_j}{p_i+p_j})=\prod_{1 \leq i <
j \leq 2N} ( \frac{p_i-p_j}{p_i+p_j}), \label{ca} \ee
we obtain from (\ref{pf1}) 
\be \tau (x,y,t)=\prod_{1 \leq i <
j \leq 2N} ( \frac{p_i-p_j}{p_i+p_j})E_1 E_2 E_3 \cdots E_{2N}. \label{fu} \ee
This gives the trivial solution.

\indent The paper is organized as follows. In section 2, one constructs the real soliton solution using the $\tau$ function and the Schur Identity. By the minor-summation formula of the Pfaffian, we can study the interactions  of solitons in the BKP  equation (\ref{bkp}) using the totally non-negative  Grassmannian. In section 3, we investigate N-soliton solutions and the corresponding self-dual $\tau$-functions. The section 4 is devoted to the Grassmannian of the Sawada-Kotera equation.  In section 5, we conclude the paper with several remarks.

\section{Multi-Line Solitons} 
In this section, one introduces the real Grassmannian (or the $ N \times M$ matrix) to construct multi-line  solitons. \\
\indent To introduce the real Grassmannian (or the resonance), we have to consider linear combination of
 $ E_i(x,y,t)=e^{p_i x+ p_i^3 t+ p_i^5 t+ \eta_i}$ .  Then each resonant solution of  BKP equation can be parametrized  by a full rank matrix, $ M \geq N$,  
\[ A= \left[\ba{cccc} a_{11}  & a_{12}  & \cdots   &  a_{1M}   \\
 a_{21} &  a_{22} & \cdots   &   a_{2M} \\
\vdots  & \vdots & \vdots & \vdots    \\
 a_{N1} & a_{N2}  & \cdots  &   a_{NM}    \ea \right] \in M_{N \times M} (\textbf{R}), \]  
where $a_{ij}$ are real constants. We notice that if  $N$ is odd, then we can extend $A$ to 
\be  \hat A= \left[\ba{ccccc} a_{11}  & a_{12}  & \cdots   &  a_{1M} & 0  \\
 a_{21} &  a_{22} & \cdots   &   a_{2M} & 0 \\
\vdots  & \vdots & \vdots & \vdots    \\
 a_{N1} & a_{N2}  & \cdots  &   a_{NM}   &0 \\
0& 0 &0 &\cdots &1  \ea \right] =A \oplus [1] \in M_{(N +1) \times (M+1)} (\textbf{R}). \label{dr} \ee 
In this case, one takes $E_{M+1}=1$. \\
\indent Let's assume that 
\[f_n(x,y,t)= a_{n1} E_1+ a_{n2} E_2+a_{n3} E_3+ \cdots +a_{nM} E_M, \quad 1\leq n \leq N.\]
If $N$ is odd, using $\hat A$, we have 
\be  f_{N+1}=E_{M+1}= 1.\label{sp} \ee
Then it can be shown that
\begin{equation} Pf( f_1,f_2, f_3, \cdots, f_N)=Pf (A W A^T), \end{equation}
where the $M \times M$ matrix $W$ is defined by the element $W_{i,j}=\frac{p_i-p_j}{p_i+p_j} E_i E_j$. Also, if $N$ is odd, using $ (\ref{sp})$ , we yield $W_{i, N+1}=E_i$ for $i=1,2,3, \cdots , N$.  Now, we have the
minor-summation formula \cite{km, is}, assuming $N$ is even,  $[M]=\{1,2,3,\cdots, M\}$, 
\begin{equation} \tau_A =Pf (A W A^T)=\sum_{J
\subset [M],\quad \sharp J=N} det (A_J) Pf(W_J), \label{sum}
\end{equation}
 where
$W_J$ denotes the $N \times N $ submatrix of $W$ and $det (A_J)$ denotes the determinant ( or minor) of  the  $N \times N $ submatrix of $A$  both obtained by picking up the rows and columns indexed by  the same index set $J$; moreover, using  the Schur identity (\ref{ca}), one yields 
\be  Pf(W_J)= Pf (E_{j_1},E_{j_2}, E_{j_3}, \cdots, E_{j_N} )= \prod_{l< m} \frac{(p_{j_l}-p_{j_m}) }{(p_{j_l}+p_{j_m})}E_{j_1} E_{j_2}E_{j_3} \cdots E_{j_N}. \label{va} \ee
It is a generalization of the $\tau$-function (\ref{fu}). We also notice that the coefficients $ det (A_J) $ of $\tau_A$   satisfies the Plucker relations, which are different from the ones obtained by Pfaffian defined on strict partitions \cite{ni}. From this formula (\ref{sum}), one can investigate the possibility of
resonance of real solitons of the BKP equation (\ref{bkp}) using the resonance theory of KP-(II) equation \cite{ko1, ko3}, such as the Y-type resonance, the O-type,   P-type interactions and T-type resonance \cite{ko6}. To obtain non-singular soliton solutions, we have to assume $ det (A_J) \geq 0$ for each index $J$. Therefore, it is is necessary to use the totally non-negative Grassmannian \cite{ko3, ko6}.  On the other hand,  to make the product in (\ref{va}) is positive, one also assumes that  
\be   p_1>p_2 >p_3> \cdots > p_N >0  \quad or \quad   p_1< p_2  <p_3< \cdots  < p_N <0.   \label{pv} \ee
\indent From the form of $\tau$-function (\ref{sum}) and (\ref{va}), the $xy$-plane is partitioned into several regions depending on the exponential parts 
\be   E_{j_1} E_{j_2}E_{j_3} \cdots E_{j_N}=e^{(p_{j_1} +p_{j_2}+\cdots+p_{j_N})x+(p_{j_1}^3 +p_{j_2}^3+\cdots+p_{j_N}^3)y  + (p_{j_1}^5 +p_{j_2}^5+\cdots+p_{j_N}^5)t +\eta_{j_1}+\eta_{j_2}+\cdots+ \eta_{j_N}} \label{ph} \ee
 in its own region. Each line soliton is obtained by the balance between adjacent regions and is localized only at the boundaries of the dominant regions. For any fixed time, suppose $ E_{i} E_{j_2}E_{j_3} \cdots E_{j_N}$ and $E_{j} E_{j_2}E_{j_3} \cdots E_{j_N} $ are adjacent regions. In general,  one has the boundary, i.e., the  line soliton $ [i,j]$-soliton. From (\ref{sum}) , (\ref{va}), and (\ref{ph}), it can be seen locally 
\bean  \tau_A  & \approx   & det (A_I) \prod_{j_l< j_m} \frac{(p_{j_l}-p_{j_m}) }{(p_{j_l}+p_{j_m})} E_i E_{j_2} E_{ j_3}    \cdots, E_{j_N} \\
& + & det (A_J)\prod_{ j_l< j_m} \frac{(p_{j_l}-p_{j_m}) }{(p_{j_l}+p_{j_m})}    E_j E_{j_2} E_{ j_3}    \cdots, E_{j_N} ,   \eean 
where $I=\{i, j_2, j_3,  \cdots , j_N\} $ and $J =\{j , j_2, j_3,  \cdots , j_N\}. $  Since $u_1=\phi_x=2 \pa_{xx} (\ln \tau)$ in (\ref{bkp}), we have the  $ [i,j]$-soliton 
\bea  u_1&=& 2 \pa_{xx} [1+ e^{(p_i-p_j)x+ (p_i^3-p_j^3)y+  (p_i^5-p_j^5)t+ \mu_i-\mu_j}] \no  \\ 
&=& \frac{(p_i-p_j)^2}{2} sech^2 \frac{(p_i-p_j)x+ (p_i^3-p_j^3)y+  (p_i^5-p_j^5)t+\mu_i-\mu_j }{2} \label{am}.
\eea
where 
\bean 
\mu_i &=& \eta_i+ \ln \left [ det (A_I) \prod_{j_l< j_m } \frac{(p_{j_l}-p_{j_m}) }{(p_{j_l}+p_{j_m})} \right ] \\
\mu_j &=&\eta_j +\ln  \left [det (A_J)\prod_{ j_l< j_m} \frac{(p_{j_l}-p_{j_m}) }{(p_{j_l}+p_{j_m})} \right ].  
\eean 

Hence the  soliton is localized along the line 
\be (p_i-p_j)x+ (p_i^3-p_j^3)y+  (p_i^5-p_j^5)t+ \mu_i-\mu_j=0. \label{pi} \ee
Also,  we notice that the amplitude is  $ \frac{(p_i-p_j)^2}{2}$ and the slope of soliton is measured in the counter-clockwise sense from the y-axis given by
\be  \tan \theta_{[i,j]}=p_i^2+p_i p_j+ p_j^2=(p_i+\frac{1}{2}p_j)^2+\frac{3}{4}p_j^2 >0 . \label{tan} \ee
It can be seen that $ 0<   \theta_{[i,j]}< \frac{\pi}{2}$. \\
Example: Let 
\[   A=\left[\ba{ccccc} 1 &0& -a & 0 & b  \\ 0 & 1 & c & 0& -d \\  0 & 0 & 0 & 1& e \ea \right ],   \]
where $a,b, c,d,e $ are positive constants and $ bc-ad \geq 0$ \cite{ko1}. Then $A$ is a totally non-negative Grassmannian. We extend A to 
\[   \hat A=\left[\ba{cccccc} 1 &0& -a & 0 & b  &0 \\ 0 & 1 & c & 0& -d &0  \\  0 & 0 & 0 & 1& e &0 \\
     0 &0& 0& 0 & 0  &1  \ea \right ].  \]
Also, 
\bean 
f_1 &= & E_1+0 E_2-aE_3+0E_4+bE_5+0E_6 , \quad  f_2= 0E_1+ E_2+cE_3+0E_4-dE_5+0E_6 \\
f_3 &= & 0E_1+0 E_2+0E_3+E_4+eE_5+0E_6, \quad f_4= E_6=1. 
\eean 
From (\ref{sum}), a direct calculation obtains 
\bean 
\tau_{\hat A} &=& Pf(f_1, f_2, f_3, f_4)=Pf(f_1, f_2, f_3, 1) \\
&=& Pf \left[\ba{cccc} 0  & W(f_1, f_2)&W(f_1, f_3)  & W(f_1, 1)  \\ W(f_2, f_1) & 0 &W(f_2, f_3)  &W(f_2, 1)  \\  W(f_3, f_1) & W(f_3, f_2) & 0 &   W(f_3, 1)  \\
     W(1, f_1) &W(1, f_2)& W(1, f_3)& 0   \ea \right ] \\
& =& 0\alpha_{123} E_1E_2E_3+ \alpha_{124}E_1E_2 E_4+ e \alpha_{125}E_1E_2 E_5+c \alpha_{134}E_1E_3 E_4 \\ &+& ce \alpha_{135}E_1E_3 E_5 
              + d \alpha_{145}E_1E_4E_5+ a \alpha_{234}E_2E_3E_4+ a e \alpha_{235}E_2E_3 E_5 \\
							&+& b \alpha_{245}E_2E_4 E_5+ (bc-ad) \alpha_{345}E_3 E_4E_5,
\eean
where 
\[ \alpha_{ijk}= \frac{(p_i-p_j) (p_i-p_k)(p_j-p_k) }{  (p_i+p_j) (p_i+p_k)(p_j+p_k), } \quad i<j<k    \]
and $ W(f_i, f_j)$ is the skew product defined in (\ref{sk}) and   $W(f_i, 1)=f_i$.  Also, the coefficient of $E_i E_j E_k$ before $\alpha_{ijk}$ comes from the determinant of the i-th, j-th and k-th columns of $ A$.  \\

\section{ N-Soliton Solutions and Dual Grassmannians }
\indent We notice that the BKP equation (\ref{bkp}) or the Hirota  equation (\ref{hr}) is invariant under the transformation $u(x,y,t) \to u(-x,-y, -t)$, as similar to the KP equation \cite{bc, bca, ko6}. Then one can investigate the self-dual $\tau$ -functions for the BKP equation. In this section, one studies the self-dual 
$\tau$-functions of   the N-soliton solutions, that is, those real line solitons  for which the sets of incoming and outgoing  $ ( |y| \to \infty)  $ asymptotic line solitons are the same.  \\
\indent There are three basic types of N-soliton: P-type, O-type and T-type. The N-soliton solutions are combinations of these three line solitons \cite{ko6}.  Notice that the "X" -shaped P-type and O-type  solitons are not resonant case, but the "box"-shaped T-type soliton is the resonant one. For the resonance of Y-type,  the P-type and O-type interactions of the BKP equation (\ref{bkp}), one can refer to \cite{yq} to obtain the numerical simulations.\\
\indent For N-soliton solutions, one considers an $ N \times 2N$ matrix. From the equation (\ref{sum}), $\tau$-function is a non-negative sum of the same exponential phase combinations  with a different sets of coefficients satisfying  the Plucker relations. Using (\ref{sum}), we see that 
\[ \tau(-x, -y, -t)= e^{-(\theta_1+\theta_2+\cdots+\theta_N)} \wt \tau(x,y,t)\], 
where $\theta_i=p_i x+ p_i^3 t+ p_i^5 t+ \eta_i, i=1,2,\cdots, N$ and 
\be \wt \tau(x,y,t)=\sum_{J \subset [2N],\quad \sharp J=N} det (A_J) \prod_{ j_l< j_m} \frac{(p_{j_l}-p_{j_m}) }{(p_{j_l}+p_{j_m})} E_{i_1} E_{i_2}E_{i_3} \cdots E_{i_N} .\label{pr} \ee
Here $\{j_1, j_2, \cdots, j_N\}$ and $\{i_1, i_2, \cdots, i_N\}$ forms a disjoint partition of $\{1,2,3,4, \cdots, 2N\}$. Now, one assumes when $ | y| \to \infty$ , the unbounded line solitons are the same, that is, the N incoming line solitons and the N  outgoing coincides. Thus, we can define a self-dual $\tau$-function \cite{bca}. Given an $N \times 2N$, irreducible, rank $N$ coefficient matrix $A$ whose $N \times N$ minors satisfies the duality conditions
\be det (A_J)=0 \quad iff \quad det (A_I)=0,  \label{du} \ee
where $J =\{ j_1, j_2, \cdots, j_N   \}$ and $ I= \{i_1, i_2, \cdots, i_N\}$ forms a disjoint partition of $\{1,2,3,4, \cdots, 2N\}$. A $\tau-$function define in (\ref{sum}) satisfies the condition (\ref{du}) is called  self-dual.  Then from (\ref{pr}), we can construct the dual  Grassmannian $B$ such that,  up to an overall factor, 
\be \wt  \tau(x,y,t)=\sum_{I  \subset [2N], \sharp I=N} det (B_I) \prod_{ i_l< i_m} \frac{(p_{i_l}-p_{i_m}) }{(p_{i_l}+p_{i_m})} E_{i_1} E_{i_2}E_{i_3} \cdots E_{i_N} ,\label{wt} \ee
where $[2N]=\{1,2,3,\cdots, 2N\}.$
It can be constructed as follows. 
\begin{itemize}

\item $N$  is even.  Suppose $A$ is written in RREF form and it's given as 
\[ A=[I_N, G]P, \]
where $I_N$ is the $N \times N$ identity matrix ,  $G$ is an  $N \times N$ matrix and $P$ is an $2N \times 2N$ permutation matrix satisfying $P^T =P^{-1}$. Here $ P^T$  is the transpose of  $P$. We define 
\[ \wt A =[ - G^T, I_N]P. \] 
We see that $A \wt A^T=0$ and consider 
\[H =\left[\ba{cc} I_N & G      \\ -G^T  & I_N \ea  \right]P . \]
Using the Laplace expansion, one yields
\be det (H)= \sum_{J\cup I =[2N]} \sigma (J,I) det  (A_J) det  (\wt A_I), \label{la} \ee
where $J={ j_1 < j_2 < j_3 < \cdots< j_N} $ and $I={ i_1 < i_2 < i_3 < \cdots< i_N} $  form a partition of $[2N]$ and $\sigma (J,I)$ is the sign function for the permutation $ \pi= (J, I)$, that is, $\sigma  (J,I)=(-1)^{\frac{N(N+1)}{2} +j_1+j_2+ \cdots +j_N}$.  On the other hand, noticing that $ P^T=P^{-1}$, 
\bea 
det H &=&  (det P)  det \left[\ba{cc} I_N & G      \\ -G^T  & I_N \ea  \right]=( det P ) det ([I_N+ GG^T]) \no \\
&=& ( det P )  det ([I_N, G]\left[\ba{c} I_N^T  \\ G^T  \ea  \right]) =(det P) det (A A^T)   \no \\
&=& (det P) \sum_{J \in  {[2N] \choose N} } (Det (A_J))^2,  \label{de} 
\eea 
where we have used the Binet-Cauchy formula in the last equality. The equations (\ref{la}) and (\ref{de}) imply that 
\be  \sigma (J,I) det (\wt A_I)=det (P) det (A_J) \label{ke}. \ee
Let's define the $2N \times 2N$ matrix  $D=diag (-1, (-1)^2, (-1)^3,  \cdots, (-1)^{2N})=diag (-1, 1, -1, 1 , \cdots , -1, 1)$ and $B= \wt A D$. Then the Binet-Cauchy formula again obtains 
\[  det  (B_I)=(-1)^{i_1+i_2+ \cdots+ i_N} det ( \wt A_I) .\] 
Consequently, it follows from (\ref{ke}) that 
\be det (P )  det ( A_J)= (-1)^{\frac{N(N+1)}{2} +\frac{2N(2N+1)}{2}}  det (B_I). \label{co} \ee
Since  $ u_1=2 (\ln \tau)_{xx}$, the equation (\ref{pr}) can be written as 
\be \wt \tau(x,y,t)=\sum_{I \subset [2N], \sharp I=N} K_J det (B_I) E_{i_1} E_{i_2}E_{i_3} \cdots E_{i_N} , \label{mo} \ee
where  $ K_J=\prod_{ j_l< j_m}  \frac{(p_{j_l}-p_{j_m}) }{(p_{j_l}+p_{j_m})}.$
Next, let's choose $\eta_i$ defined in the phase $\theta_i$ as 
\[\eta_i= \sum_{r \neq i} \ln   \frac{|(p_r-p_i) |}{|(p_r+p_i)|}, \quad i, r =1,2,3, \cdots, 2N . \]
Then we have 
\be  e^{ \sum_{m=1}^N \eta_{i_m}}= \prod_{m=1}^N \prod_{r \neq i_m}  \frac{|(p_r-p_{i_m}) |}{|(p_r+p_{i_m})|}      =K_J^{-1} K_I K_{[2N]}\quad   \mbox{with} \quad I \cup J=[2N],            \label{ff} \ee
where $K_{[2N]}= \prod_{i <j}  \frac{(p_i-p_j) }{(p_i+p_j)},$ an overall factor. Plugging (\ref{ff}) into (\ref{mo}), one has the equation (\ref{wt}). 
\item $N$  is odd. In this case, for the Pfaffian structure (\ref{sum}),   one considers the $(N+1) \times (2N+1)$  matrix $\hat A= A\oplus [1]$, as is defined in (\ref{dr}). Define the $(N+1) \times (2N+1)$  matrix  $\hat {\wt A} =[ - G^T, I_N]P  \oplus [1]$ and we see that the $(N+1) \times (N+1)$  matrix $ \hat A \hat {\wt A}^T = diag (0,0,0,\cdots, 1)$.  Also, one defines the $(2N+1) \times (2N+1) $ matrix 
\[\hat H= H \oplus [1]=\left[\ba{cc} I_N & G      \\ -G^T  & I_N \ea  \right]P \oplus [1] ,\]
which has the same determinant of $H$. The matrix $B$ becomes the $(N+1) \times (2N+1)$ matrix $\hat B= B\oplus [1]$.  Hence 
\[ det (\hat B_{\hat I})=det  (B_I)=(-1)^{i_1+i_2+ \cdots+ i_N} det ( \wt A_I), \]
where $\hat I= I \cup \{2N+1\}=\{i_1, i_2, i_3,  \cdots,  i_N, 2N+1\} $.  Then a similar consideration also yields the equation (\ref{wt}) when $N$ is odd. 
\end{itemize}
Finally, one remarks that if $B=\wt A D=A$, then we have 
\[A D A^T=A DD^T \wt{A}^T= A \wt{A}^T=0. \]
From this, one defines the Orthogonal Grassmannian \cite{gp}
\be  OG(N, 2N)=\{ A \in R^{N\times 2N} |  A D A^T=0\} , \label{og} \ee
where $D=diag (-1, 1, -1, 1 , \cdots , -1, 1)$. The condition of  self-dual condition (\ref{du}) is weaker than Orthogonal Grassmannian. The dimension of $  OG(N, 2N) $ is $ \frac{N(N-1)}{2}$  and it has the important property
\[ det (A_J)= det (A_I), \]
where $J={ j_1 < j_2 < j_3 < \cdots< j_N} $ and $I={ i_1 < i_2 < i_3 < \cdots< i_N} $  form a partition of $[2N]$. 
For any $A \in OG(N, 2N)$, it can be seen that, up to an overall factor, 
\[ \tau_A (x,y,t; p_1,p_2, \cdots, p_{2N})=\tau_A (-x,-y,-t; p_1,p_2, \cdots, p_{2N})= \tau_A (x,y,t; -p_1, -p_2, \cdots, -p_{2N}). \]
Example (T-type Soliton): The totally non-negative Grassmannian is the $ 2\times 4$ matrix \cite{ko1}
\[A_T=\left[\ba{cccc} 1 & 0 &-c &-d       \\ 0& 1 & a & b  \ea  \right] , \]
where $ a,b, c,d >0$ and $ad-bc >0$. In this case, $P=I_2$ and 
\[\hat A_T =\left[\ba{cccc} c & -a &1  &0      \\ d & -b & 0 & 1  \ea  \right] . \]
Since 
\[f_1 =  E_1-cE_3-dE_4,  \quad f_2 =   E_2+aE_3+bE_4,\]
one has 
\bea
\tau_{A_T} &=& Pf(f_1, f_2) = \frac{p_1-p_2}{p_1+p_2} E_1E_2 +a \frac{p_1-p_3}{p_1+p_3} E_1E_3+b\frac{p_1-p_4}{p_1+p_4} E_1E_4 \no \\ 
&+ & c  \frac{p_2-p_3}{p_2+p_3} E_2E_3+d  \frac{p_2-p_4}{p_2+p_4} E_2E_4 +  (ad-bc)   \frac{p_3-p_4}{p_3+p_4} E_3E_4.
\eea
The numerical simulation shows that,  for  $ y>>0$, the asymptotic line solitons from left to right are the $[1, 3]$ and $[2, 4]$ solitons  due to the conditions (\ref{pv}) and (\ref{tan})  when $t >0$;  for  $ y< <0$, the asymptotic line solitons from left to right are the $[2, 4]$ and $[1, 3]$ solitons. The $[1,4]$ soliton has the highest amplitude and there is a box formed by the four line solitons in the $xy$-plane  to describe the resonance for any time.  As for $t <0$, the figure is obtained by reflecting the figure of $t>0$ with respect to  the origin $(0,0)$.   To obtain the Orthogonal Grassmannian (\ref{og}), using $D=diag(-1, 1, -1, 1)$, we have the conditions: $a=d, b=c $ and $a^2-b^2=1$.

\section{ The Grassmannian of Sawada-Kotera Equation}
In this section, we investigate the Grassmannian of the real soliton  of the Sawada-Kotera Equation (SKE) \cite{kk}, which is a fifth-order non-linear differential equation.   It is obtained from (\ref{bkp}) and is independent of $y$. The SKE is written as the equation (in the potential form)
\be 
(9 \phi_t+\phi_{xxxxx}+15 \phi_{x}\phi_{xxx}+15\phi_x^3)=0,  \label{sk}
\ee
which has the Hirota Equation by (\ref{hr})   
\be D_x(D_x^5+9 D_t) \tau \circ \tau=0.  \label{sh} \ee
The SKE (\ref{sk}) has the Lax representation \cite{mv} 
\[   [\pa_x^3 + 3 \phi_x\pa_x] \psi = \la \psi   \]
 \[ [ \pa_t + (9\phi_{xx} -9 \la ) \pa_x^2+ (9\phi_x^2-3 \phi_{xxx}) \pa_x] \psi = 18 \lambda \phi_x \psi.  \]
Let  $\phi=0$ and $ \lambda=\lambda_i$ in (\ref{la}), and  it results that $ E_i = \psi_i(\lambda_i ) $  satisﬁes the linear equations
\be E_{ixxx} - \la_i E_i = 0,\quad E_{it }- 9\lambda_i E_{ixx}  = 0,  \label{ie} \ee
with the general solution
\be E_i = A_i e^{q_i x + 9q_i^5 t}+ B_i e^{r_i x + 9r_i^5 t}+  C_i e^{s_i x + 9s_i^5t}, \label{so}\ee
where $A_i, B_i, C_i $ are arbitrary complex constants, and $q_i , r_ i, s_i $ are the three different cubic roots of complex number $\lambda_i$, that is, 
\be  q_i=(\la_i)^{1/3},  \qd r_i=\omega (\la_i)^{1/3}, \qd  s_i=\om^2 (\la_i)^{1/3}, \qd \om^2+\om+1=0.\label{qr}
 \ee
Defining $r_i-q_i=p_i$,  we have $9 (r_i^5-q_i^5)= -p_i^5$.  Also, we choose a complex number $\lambda_i$ appropriately such that $p_i$ is a real number. Then one sets $C_i=0$. \\
\indent To obtain $N$-solitons solution, one takes $\la_1, \la_2, \cdots, \la_N ,  q_1, q_2, \cdots , q_N$  and $ r_1, r_2, \cdots, r_N$ in (\ref{qr}) such that $r_i-q_i=p_i , i=1,2, \cdots, N$ are real numbers.  If $N$ is odd, then we take $f_{N+1}=1 $ by (\ref{dr}).  We consider the P-type soliton (p.110, \cite{ko6})
\be f_i=A_i e^{ q_i x+ 9q_i^5 t}+(-1)^{N-i} B_i e^{  r_i x+ 9 r_i^5 t}=A_i \theta_i+ (-1)^{N-i} B_i \rho_i, \quad i=1,2, \cdots, N\label{le},  \ee
where we define $\theta_i=e^{ q_i x+ 9q_i^5 t}$ and $\rho_i= e^{  r_i x+ 9 r_i^5 t}$ and the Grassmannian  corresponding to the P-type (non-resonant case) is the $N \times 2N$ matrix : 
 \be  A_P=\left[\ba{cccccccc} A_1 & 0 & \cdots  & 0  &0 &\cdots &  0 & (-1)^{N-1}B_1\\ 
 0 & A_2 & \cdots  & 0  &0 &\cdots & (-1)^{N-2} B_2 & 0 \\
\vdots  & \ddots & \ddots  & \vdots &\vdots &\ddots &  \vdots & \vdots\\
 0 & 0 & 0  & A_N  &B_N & 0&  \cdots  & 0 \ea
 \right]. \label{ma} \ee  
Then 
 \[ \left[\ba{c} f_1 \\ f_2 \\ f_3 \\  f_4  \\  \vdots  \\  \vdots \\   \vdots \\ f_N  \ea
 \right]= A_P  \left[\ba{c} \theta_1\\  \theta_2 \\  \vdots  \\  \theta_N  \\ \rho_N \\ \vdots   \\ \rho_2  \\ \rho_1  \ea \right]      \]
Without loss of generality, we assume $N$ is even by the section 3. Then from the Schur identity (\ref{ca}) and (\ref{sum})  the Pfaffian 
\bea  
\tau_{A_P}&=&Pf  (f_1, f_2, f_3, \cdots,  f_N) = A_1 A_2 \cdots A_N \prod_{i<j \leq N}\frac{q_i-q_j}{q_i+q_j}   \theta_1 \theta_2 \cdots \theta_N  \no \\
&+& \sum_{I \cup J= [N]}  A_1 A_2 \cdots   A_{j_1-1} B_{j_1} A_{j_1+1} \cdots A_{j_2-1} B_{j_2}A_{j_2+1} \cdots  A_{j_n-1}B_{j_n}  A_{j_n+1}, \cdots  A_N  \no \\
& \times &  \chi (q_1, q_2, \cdots, q_{j_1-1} , r_{j_1}, q_{j_1+1}, \cdots,  q_{j_2-1}, r_{j_2} ,  q_{j_2+1},  \cdots,  q_{j_n-1} , r_{j_n},  q_{j_n+1}, \cdots , q_N )  \no \\
& \times &  \theta_1 \theta_2 \cdots   \theta_{j_1-1} \rho_{j_1} \theta_{j_1+1} \cdots \theta_{j_2-1} \rho_{j_2}\theta_{j_2+1} \cdots  \theta_{j_n-1}\rho_{j_n}  \theta_{j_n+1} \cdots  \theta _N \no \\
&+&  B_1 B_2 \cdots B_N\prod_{N \geq \alpha > \beta}\frac{r_{\alpha}-r_{\beta} }{ r_{\alpha}+r_{\beta} }   \rho_1 \rho_2 \cdots \rho_N,   \label{su}
\eea
where $I=\{1,2, \cdots , j_1-1, j_1+1, \cdots,  j_2-1, j_2+1,  \cdots,  j_n-1, j_n+1, \cdots , N \}$ and $J=\{j_1, j_2, \cdots, j_n\}$  form a partition of $[N]=\{1,2,3,, \cdots , N\}$. Also, 
\bea   &&  \chi (q_1, q_2, \cdots, q_{j_1-1} , r_{j_1}, q_{j_1+1}, \cdots,  q_{j_2-1}, r_{j_2} ,  q_{j_2+1},  \cdots,  q_{j_n-1} , r_{j_n},  q_{j_n+1}, \cdots , q_N ) \no \\
&=& \prod_{i<j, \,   i, j \in I}\frac{q_i-q_j}{q_i+q_j} \prod_{\alpha > \beta , \,   \alpha, \beta  \in J}\frac{r_{\alpha}-r_{\beta} }{ r_{\alpha}+r_{\beta} }\prod_{  \gamma \in I, \epsilon  \in  J}\frac{q_{\gamma}-r_{\epsilon} }{ q_{\gamma}+r_{\epsilon} }. \label{pd} 
 \eea
There are $2^N$  terms in the expansion (\ref{su}).  Since $u_1=2\pa_{xx} \ln \tau_{A_P} $, we factor out the gauge function $\theta_1 \theta_2 \cdots \theta_N$ and the overall coefficient $\chi_q= \chi(q_1, q_2, \cdots, q_N)=\prod_{i<j } \frac{q_i-q_j}{q_i+q_j} $.  Then  the $\tau$-function defined in (\ref{su}) is equivalent to, using (\ref{qr}), 
\bea  
 \tau_{A_P}  &\cong & A_1 A_2 \cdots A_N  \no \\
&+& \sum_{I \cup J= [N]}  A_1 A_2 \cdots   A_{j_1-1} B_{j_1} A_{j_1+1} \cdots A_{j_2-1} B_{j_2}A_{j_2+1} \cdots  A_{j_n-1}B_{j_n}  A_{j_n+1}, \cdots  A_N  \no \\
& \times & \chi_q^{-1}  \chi (q_1, q_2, \cdots, q_{j_1-1} , r_{j_1}, q_{j_1+1}, \cdots,  q_{j_2-1}, r_{j_2} ,  q_{j_2+1},  \cdots,  q_{j_n-1} , r_{j_n},  q_{j_n+1}, \cdots , q_N )  \no \\
& \times &  \prod_{j=1}^n e^{p_j x - 9 p_j^5 t}+ B_1 B_2 \cdots B_N \chi_q^{-1} \prod_{N \geq \alpha > \beta}\frac{r_{\alpha}-r_{\beta} }{ r_{\alpha}+r_{\beta} }  \prod_{j=1}^N e^{p_j x - 9 p_j^5 t} , \label{red}
\eea
where $\cong$ means equivalence. \\
\indent Next, letting $A_1=A_2=\cdots =A_N=1$, the equation (\ref{red}) becomes 
\bea  
 \tau_{A_P}  &\cong & 1+ \sum_{I \cup J= [N]}   B_{j_1}  B_{j_2}\cdots B_{j_n}    \no \\
& \times & \chi_q^{-1}  \chi (q_1, q_2, \cdots, q_{j_1-1} , r_{j_1}, q_{j_1+1}, \cdots,  q_{j_2-1}, r_{j_2} ,  q_{j_2+1},  \cdots,  q_{j_n-1} , r_{j_n},  q_{j_n+1}, \cdots , q_N )  \no \\
& \times &  \prod_{j=1}^n e^{p_j x - 9 p_j^5 t}+ B_1 B_2 \cdots B_N \chi_q^{-1} \prod_{N \geq \alpha > \beta}\frac{r_{\alpha}-r_{\beta} }{ r_{\alpha}+r_{\beta} }  \prod_{j=1}^N e^{p_j x - 9 p_j^5 t}. \label{re}
\eea
Now, one defines the variables, $j=1,2, \cdots, N$, 
\be  \Omega_j=\frac{\chi( q_1, q_2, \cdots, q_{j-1}, r_{j}, q_{j+1}, \cdots, q_N)}{\chi_q}=\frac{\prod_{m=1, m \neq j}^N \frac{q_{m}-r_{j} }{ q_{m}+r_{j} }}{  \prod_{m <j} \frac{q_{m}-q_{j} }{ q_{m}+q_{j} } \prod_{m >j} \frac{q_{j}-q_{m} }{ q_{j}+q_{m} } }. \label{om} \ee
Then we have the relations
\bea 
 && \chi (q_1, q_2, \cdots, q_{j_1-1} , r_{j_1}, q_{j_1+1}, \cdots,  q_{j_2-1}, r_{j_2} ,  q_{j_2+1},  \cdots,  q_{j_n-1} , r_{j_n},  q_{j_n+1}, \cdots , q_N ) \no \\
&=& \chi_q \Omega_{j_1}  \Omega_{j_2} \cdots  \Omega_{j_n} \left( \prod_{\alpha < \beta} P_{j_{\alpha} j_{\beta}}\right ), \label {un}
\eea
and 
\be
 \chi (r_1, r_2 , \cdots, r_N)= \prod_{N \geq \alpha > \beta}\frac{r_{\alpha}-r_{\beta} }{ r_{\alpha}+r_{\beta} } =\chi_q \Omega_{1}  \Omega_{2} \cdots  \Omega_{N} \left( \prod_{\alpha < \beta} P_{\alpha \beta}\right ), \label {un2}
\ee
where 
\bea
P_{j_{\alpha} j_{\beta}}&=& \frac{\chi (q_1, q_2, \cdots, q_{j_{\alpha}-1} , r_{j_{\alpha}}, q_{j_{\alpha}+1}, \cdots,  q_{j_{\beta}-1}, r_{j_{\beta}} ,  q_{j_{\beta}+1},  \cdots , q_N ) }{\chi_q \Omega_{j_{\alpha} }   \Omega_{j_{\beta} } }\no \\
&=& \frac{(q_{j_{\alpha}}+ r_{j_{\beta}})  (q_{j_{\beta}}+ r_{j_{\alpha}})  (q_{j_{\alpha}}- q_{j_{\beta}})       (r_{j_{\beta}}- r_{j_{\alpha}})  } {(q_{j_{\alpha}}- r_{j_{\beta}})  (q_{j_{\beta}}- r_{j_{\alpha}})  (q_{j_{\alpha}}+ q_{j_{\beta}})       (r_{j_{\beta}}+ r_{j_{\alpha}})} \no \\
&=& \frac{   (p_{j_{\alpha}}- p_{j_{\beta}})^2 ( p_{j_{\alpha}}^2+  p_{j_{\alpha}} p_{j_{\beta}}+    p_{j_{\beta}}^2 )  }{ (p_{j_{\alpha}}+ p_{j_{\beta}})^2 ( p_{j_{\alpha}}^2-  p_{j_{\alpha}} p_{j_{\beta}}+   p_{j_{\beta}}^2 )  }= P_{j_{\beta} j_{\alpha}}.   \label{pi}
\eea
For example, 
\[\chi (q_1, r_2 , r_3, r_4, q_5, r_6)=\chi_q \Omega_2 \Omega_3\Omega_4 \Omega_6P_{23} P_{24}P_{26}
   P_{34} P_{36}    P_{46}.  \]
Here we notice that all the $P_{j_{\alpha} j_{\beta}}$ could be positive real numbers. The proof of equation (\ref{un}) (or  (\ref{un2})) is given in the appendix. \\
\indent Letting $b_i=B_i \Omega_i >0 , i=1,2, \cdots, N$, using the relations (\ref{un}) and (\ref{un2}), the $\tau$-function in (\ref{re}) has the standard form of N-solitons obtained by the vertex operator  \cite{dj, hi}
\[\tau_{A_P}  \cong  \sum \exp \left [\sum_{i=1}^N \zeta_i \Upsilon_i + \sum_{i<j}^{(N)} \Theta_{ij} \zeta_i \zeta_j \right ] ,\]
where $\sum$ denotes the summation over all possible combinations of $ \zeta_i=0,1$ , where  $ i=1,2, \cdots , N$, and $\sum_{i<j}^{(N)}$ is the sum over all pairs $i,j$ chose from ${1,2,3, \cdots, N}$ ; moreover, 
\[\Upsilon_i=p_i x - 9 p_i^5 t+ \ln b_i,  \quad \Theta_{ij}= \ln P_{ij}.\]
Finally, if we redefine in (\ref{le}) 
\[f_i=A_i e^{ q_i x+ 9q_i^5 t}+(-1)^{N-i} \frac{b_i} {\Omega_i} e^{  r_i x+ 9 r_i^5 t}, \]
 then (\ref{ma}) has the form 
\be  A_P=\left[\ba{cccccccc} 1 & 0 & \cdots  & 0  &0 &\cdots &  0 & (-1)^{N-1}b_1\\ 
 0 & 1 & \cdots  & 0  &0 &\cdots & (-1)^{N-2} b_2 & 0 \\
\vdots  & \ddots & \ddots  & \vdots &\vdots &\ddots &  \vdots & \vdots\\
 0 & 0 & 0  & 1  &b_N & 0&  \cdots  & 0 \ea
 \right],  \no  \ee
where $b_i, i=1,2,\cdots N $  are  positive numbers. 
\section{Concluding Remarks}
\indent In this article, one investigates the interactions of real line solitons  of the BKP  equation (\ref{bkp})  using the real Grassmannian.  We solve the linear equations (\ref{li}) and expresses their solutions as linear combinations of elementary solutions $E_i(x,y,t,) $ via the totally non-negative Grassmannians. To obtain non-singular solutions, we also need the condition (\ref{pv}).  By the minor-summation formula and the Schur Identity, one can write the $\tau$-function as a linear combination of pfaffians (\ref{sum}), whose coefficients satisfy the Plucker relations. Thus,  the resonant structure of the BKP equation can be investigated. To study the N-soliton solutions, that is, those real line solitons having the same incoming and outgoing  $ ( |y| \to \infty)  $ asymptotic line solitons, the self-dual $\tau$-functions of the BKP equation are described. In particular, the Grassmannian of P-type soliton of the Sawada-Kotera is carefully researched and a relation with the formula constructed by the vertex operator acting on the $\tau$-function defined on the Sato orthogonal Grassmannian manifold \cite{da} is proved. \\
 \indent A number of  important questions remain for further development. First off, the $\tau$-function of CKP equation obtained form the reduction of C-type in the KP hierarchy is studied in \cite{lu} and its Backlund transformation is obtained in \cite{mv} via a expression of Gram determinant structure. The interactions of real line solitons might be interesting.  Also, in \cite{kr}, the BKP hierarchy can be used as integrable deformation on the Schrodinger operator,  and its Lax formulation is the Manakov L-A-B triples. The $\tau$-function structure is investigated in \cite{hi}. The resonant theory could be further studied.  Finally,
the hypergeometric $\tau$-functions  \cite {ni} defined on all the subset of strict partitions of any finite non-negative set  are  different from the section 2.  These coefficients are the hypergeometric functions. Their corresponding line solitions and  interactions are worth exploring. Such studies are deferred to future publications.

\appendixpage
We prove the equation (\ref{un}). \\
 \indent One considers the quotient 
\be \frac{ \chi (q_1, q_2, \cdots, q_{j_1-1} , r_{j_1}, q_{j_1+1}, \cdots,  q_{j_2-1}, r_{j_2} ,  q_{j_2+1},  \cdots,  q_{j_n-1} , r_{j_n},  q_{j_n+1}, \cdots , q_N )} {\Omega_{j_1}  \Omega_{j_2} \cdots  \Omega_{j_n} } \label{po} \ee
Using the equation (\ref{pd}) and the definition $\Omega_{j_k}$ defined in (\ref{om}), one yields
\bean 
 (\ref{po}) &=& \chi_q \prod_{\alpha < \beta} \frac{ (q_{j_{\alpha}}- q_{j_{\beta}})} {(q_{j_{\alpha}}+ q_{j_{\beta}})}  \frac{(r_{j_{\beta}}- r_{j_{\alpha}})}  {  (r_{j_{\beta}}+ r_{j_{\alpha}})} \frac{ (q_{j_{\alpha}}+ r_{j_{\beta}}) (q_{j_{\beta}}+ r_{j_{\alpha}}) } {  (q_{j_{\alpha}}- r_{j_{\beta}})  (q_{j_{\beta}}- r_{j_{\alpha}})  } \\
&=&  \chi_q  \prod_{\alpha < \beta} P_{j_{\alpha} j_{\beta}}.\eean 
In the first equality, $\frac{ (q_{j_{\alpha}}- q_{j_{\beta}})}{(q_{j_{\alpha}}+ q_{j_{\beta}})}$ comes from both $\Omega_{j_{\alpha}}$ and $\Omega_{j_{\beta}}$, one of them  putted into $\chi_q$. Also, $\frac{(r_{j_{\beta}}- r_{j_{\alpha}})}  {  (r_{j_{\beta}}+ r_{j_{\alpha}})}$ comes from the numerator of (\ref{po});moreover,   $    \frac{   (q_{j_{\beta}}+ r_{j_{\alpha}})  }{ (q_{j_{\beta}}- r_{j_{\alpha}}) }  $ and $  \frac{ (q_{j_{\alpha}}+ r_{j_{\beta}})}{   (q_{j_{\alpha}}- r_{j_{\beta}}) } $  are obtained from $\Omega_{j_{\alpha}}$ and $\Omega_{j_{\beta}}$ respectively. The second equality is obtained from the definition $P_{j_{\alpha} j_{\beta}} $ defined in (\ref{pi}).\\
\indent The relation (\ref{un2}) can be proved similarly.

\subsection*{Acknowledgments}
The author is grateful to Prof. Y. Kodama  for valuable discussions.  This work is
supported in part by the National Science and Technology Council of Taiwan under
Grant No. NSC 110-2115-M-606-001.

\end{document}